\documentclass[aps,showpacs,twocolumn,superscriptaddress]{revtex4-1}

\usepackage{graphicx}
\usepackage{verbatim}
\usepackage{tabularx}
\usepackage{color}
\usepackage{float}
\usepackage{amsmath}
\usepackage{mathtools}
\usepackage{empheq}
\newcommand{\be}{\begin{equation}}
\newcommand{\ee}{\end{equation}}
\newcommand{\bea}{\begin{eqnarray}}
\newcommand{\eea}{\end{eqnarray}}
\usepackage{dcolumn}
\usepackage{hyperref}
\usepackage{bm}
\usepackage{epsf}
\usepackage{subfigure}
\usepackage{epstopdf}%
\usepackage{amsfonts}
\usepackage{braket}
\usepackage{cancel}

\begin{document}

\title {Anomalous electronic shot noise in resonant tunneling junctions} 
\author{Anqi Mu and Dvira Segal}
\affiliation{Department of Chemistry and
Centre for Quantum Information and Quantum Control,
University of Toronto, 80 Saint George St., Toronto, Ontario, Canada M5S 3H6}

\date{\today}

\begin{abstract}
We study the behavior of shot noise in resonant tunneling junctions far from equilibrium.
Quantum-coherent elastic charge transport can be characterized by a transmission function, that is 
the probability for an incoming electron at a given energy to tunnel through a potential barrier. 
In systems such as quantum point contacts, electronic shot noise is oftentimes calculated based on a constant 
(energy independent) transmission probability, a good
approximation at low temperatures and under a small bias voltage.
Here, we generalize these investigations to far from equilibrium settings 
by evaluating the contributions of electronic resonances to the electronic current noise.
Our study extends canonical expressions for the voltage-activated shot noise
and the recently discovered delta-T noise to the far from equilibrium regime,
when a high bias voltage or a temperature difference is applied.
In particular, when the Fermi energy is located on the shoulder of a broad resonance, 
we arrive at a formula for the shot noise revealing anomalous-nonlinear behavior at high bias voltage.
\end{abstract}

\maketitle

\section{Introduction}
\label{Sintro}

Noise in electronic signals is typically undesired, yet it can be a source of information on the 
conducting system, by exposing effects concealed in the time-averaged electric current \cite{Nazarov,Reznikov-Rev,Buttiker}.
Shot noise measurements at the mesoscale and nanoscale reveal
 the fractional charge of quasiparticles in many-body systems \cite{Heiblum95}, 
 contributions of different channels to the transport \cite{Ruit99,Ruit00,Ruitenbeek06,Natelson,Scheer}, 
the crossover from ballistic to diffusive mechanism \cite{NatelsonPRB},
the valence orbital structure of the contact \cite{oren13,Vardimon16},
 activation of vibrations in molecular conducting  junctions  \cite{Tal08,orenACS,Cho},
and the onset of spin-polarized transport \cite{Berndt}.

For a stationary process, the power spectrum (or spectral density) of the current noise is defined as
\bea
S(\omega)= 2\int_{-\infty}^{\infty} d\tau e^{i\omega \tau} C(\tau),
\eea
with the time averaged autocorrelation function $C(\tau)=\langle I(\tau) I(0)\rangle -\langle I(\tau)\rangle^2$. 
Here, $I(\tau)$ is the current measured at time $\tau$.
The white noise (flat power spectrum) component \cite{1overf} of measured current fluctuations 
includes different contributions \cite{Reznikov-Rev,Buttiker}.
The thermal motion of charge carriers in electronic conductors is responsible for
the Johnson-Nyquist (thermal) noise \cite{Johnson,Nyqusit}, 
which is proportional to the temperature and the linear response electronic conductance. 
When a voltage bias is applied across the conductor, voltage-induced shot noise is activated,
and it dominates over the thermal noise at high bias and low temperatures.
Furthermore, temperature differences across channels activate an additional white-noise contribution,
the recently measured  `delta-T' noise, which is quadratic in the temperature difference \cite{Oren}. 

Recent measurements of shot noise in quantum point contacts of gold 
revealed a nonlinear (termed `anomalous') noise-voltage behavior at high bias and low temperatures \cite{Tewari}.
These observations, as well as other measurements at elevated temperatures \cite{Natelson14,Natelson16}
are anomalous in the sense that they do not follow the standard theoretical prediction 
[see  Eqs. (\ref{eq:shot}) and (\ref{eq:shotV}) below].
We recall that to derive the standard formulae,
the transmission probability of electrons to cross the constriction is assumed to be a constant, evaluated
at the equilibrium Fermi energy \cite{Buttiker}.

Different mechanisms were suggested to explain observations of anomalous shot noise. 
For example, it may arise due to local heating of electrons \cite{Natelson14}, 
interference effects due to scatterings with impurities located at the electrodes at the
vicinity of the point contact \cite{Tewari}, electron-electron and electron-phonon inelastic effects \cite{Natelson16}.
Correspondingly, recent theoretical works focused on the behavior of shot noise while taking into account
electron \cite{diventra,Belzig,Galperin17}, and spin  \cite{Fransson} correlations, as well as
electron-phonon inelastic effects \cite{Nitzan06,Ora,Novotny,Asai}.
Approximately, the impact of such processes 
can be captured within the elastic transport theory by using a voltage, 
and/or temperature dependent transmission function \cite{Natelson14,Tewari}.
Nevertheless, a consistent explanation to the variety of experimental observations of anomalous shot noise 
is still missing, even for the well-studied metallic (gold) break junction setup
 \cite{Natelson14,Natelson16, Tewari}.

\begin{figure*}
\hspace{-2mm}\includegraphics[scale=0.63]{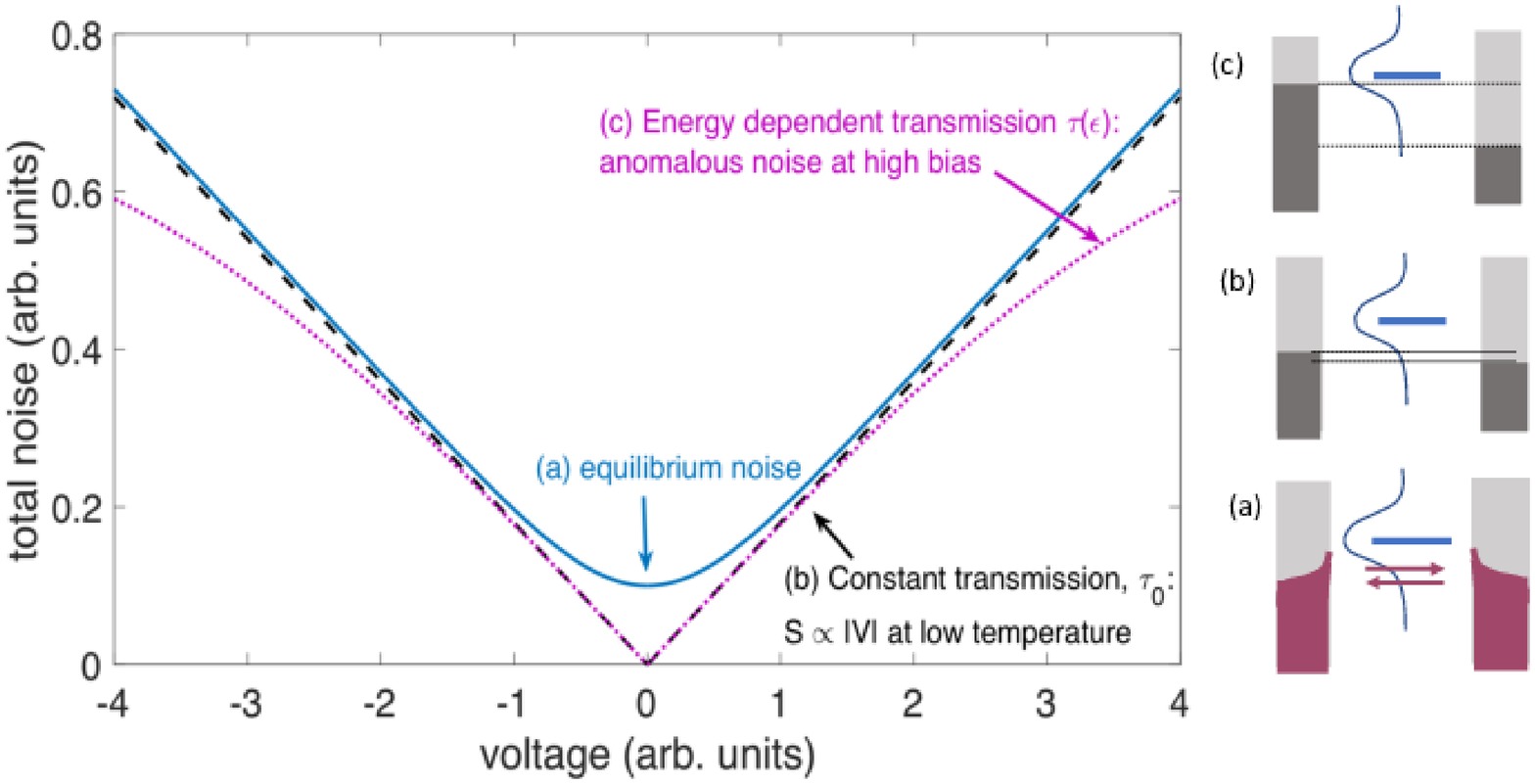}
\caption{Schematic representation of the different analytical results for the current noise as a function of voltage. 
Under the assumption of a constant transmission, the current noise obeys Eq. (\ref{eq:shot}), 
illustrated here at high (full) and low (dashed)  temperatures.
Taking into account the variation of the transmission function with energy we arrive at Eq.
(\ref{eq:SmuN}), simulated here at low temperature (dotted).
Panels (a)-(c) depict tunnel junctions with a single impurity level bridging 
two metal electrodes (a) at thermal equilibrium, (b) under low bias voltage at zero temperature,
and (c) under high bias voltage at zero temperature.
The Lorentzian lineshape illustrates the broadening of the level (resonance).
Horizontal dashed lines mark the bias window relevant for charge transport.
The behavior of the noise in the three different cases is marked on the noise-voltage curve.
}
\label{Fig0}
\end{figure*}

The objective of our work is to revisit the (seemingly solved) problem of electronic shot noise in 
quantum coherent nanostructures,
and study the behavior of shot noise in far from equilibrium situations, 
that is at high bias voltage or under large temperature differences.
Recent studies have put forward complex mechanisms for explaining observations of anomalous shot noise,
e.g. relying on many body effects. In contrast, here we carefully examine the analytically tractable 
problem of elastic conduction in resonant tunneling junctions and bring to light an overlooked,
anomalous shot noise term, which is building up far from equilibrium.
We emphasize that we restrict our analysis to the flat noise spectrum component.  

When electron transport is quantum coherent, quantum chemistry computations can be readily performed to 
take into account the rich electronic structure of the conducting channel (atoms or molecules),
as was recently done in Ref. \cite{Chen18}.
However, our goal here is to provide a more general understanding of shot noise based on analytic calculations.
Specifically, we aim to to derive formulae for the nonequilibrium zero frequency noise, which would
be useful for experimentalists in their efforts to pinpoint the origins of observed anomalous noise.

We consider resonant tunneling junctions under bias voltage or a temperature difference
and focus on the role of the transmission resonance on transport. Our central achievements are:
%
(i) Under bias voltage, we derive a closed-form expression for the shot noise---valid for broad resonances---
exposing the nonlinearity of the shot noise with bias at low temperature. 
In Fig. \ref{Fig0} we sketch the behavior of shot noise in quantum coherent conductors under low or high bias voltage,
illustrating our results.
(ii) For junctions under a temperature difference,
we show that the  $(\Delta T)^2$  scaling of the shot noise, as observed in Ref. \cite{Oren} is robust and it survives 
even when the the transmission function depends on energy. 

The paper is organized as follows. 
In Sec. \ref{secT}, we review the standard (`normal') shot noise expressions for the current noise,
received under the assumption of a constant transmission probability.
In Sec. \ref{sec-mainR}, we go beyond this assumption and present analytic results for the current noise; 
details are left to Appendices A-E.
We exemplify our derivations with simulations in Sec. \ref{sec-simul}, and conclude in Sec. \ref{sec-summ}.
 

\vspace{4mm}
\section{Standard results: constant transmission probability}
\label{secT}

We consider coherent, elastic transport of electrons in a two-terminal junction.
The metals $L$ and $R$ include collections of noninteracting electrons with occupation numbers following 
the grand canonical ensemble; the Fermi function
$f(\epsilon,\mu_{\nu},T_{\nu}) ={\frac{1}{e^{\beta_{\nu} (\epsilon-\mu_{\nu})}+1}}$
is evaluated at the chemical potential $\mu_{\nu}$ and temperature $T_{\nu}$ with the inverse temperature $\beta_{\nu}=1/k_BT_{\nu}$;
$\nu=L,R$.
Below we denote by $\mu$ the equilibrium Fermi energy. Ignoring decoherence and inelastic processes within the constriction,
the average current is given by the Landauer formula with the transmission $\tau(\epsilon)$,
\bea
\langle I\rangle = \frac{2e}{h} \int_{-\infty}^{\infty}d\epsilon\tau(\epsilon)\left[f(\epsilon, \mu_L, T_L) - f(\epsilon, \mu_R, T_R)\right].
\label{eq:Landauer}
\eea
The corresponding zero frequency power spectrum of the noise is given by \cite{Buttiker}
\begin{widetext}
\bea
S&=&S_1+S_2
\nonumber \\
S_1&=&\frac{4e^2}{h}\int_{-\infty}^{\infty}d\epsilon\{f(\epsilon, \mu_L, T_L)[1-f(\epsilon, \mu_L, T_L)]+f(\epsilon, \mu_R, T_R)[1-f(\epsilon, \mu_R, T_R)]\}\tau^2(\epsilon), \nonumber \\ 
S_2&=&\frac{4e^2}{h}\int_{-\infty}^{\infty}d\epsilon\{f(\epsilon, \mu_R, T_R)[1-f(\epsilon, \mu_L, T_L)]+f(\epsilon, \mu_L, T_L)[1-f(\epsilon, \mu_R, T_R)]\}\tau(\epsilon)[1-\tau(\epsilon)]. 
\label{eq:S12}
\eea
\end{widetext}
For simplicity, we omit the reference to (zero) frequency. 
In this partition of the total noise, $S_1$ includes additive terms in the left and right metals
while $S_2$ collects transport processes from one terminal to the other. 
The transmission function $\tau(\epsilon)$ is energy dependent; voltage and temperature dependency, rooted in many-body effects, 
are sometimes phenomenologically introduced into the transmission function---though not in our work.
%

Let us now review the standard, `normal' shot noise expressions, which are 
used to fit experimental observations.
Equations (\ref{eq:Landauer})-(\ref{eq:S12}) can be simplified if $\tau(\epsilon)$ is assumed a constant.
This assumption is justified at low bias voltage and under a small temperature difference.
Then, the width of resonances (responsible for charge transport through the system) is considerable
relative to the bias window, thus the transmission function can be approximated by its 
(fixed) value at the Fermi energy, see Fig. \ref{Fig0}(b).
Making this critical assumption, the averaged current under a finite voltage 
reduces to $\langle I\rangle =\frac{2e}{h}\Delta\mu\sum_i \tau_i$, with
the power noise \cite{Buttiker,Lesovik} 
%
\bea
S_{\Delta \mu}^T&=&4k_BTG_0\sum_{i}\tau_i^2 
\nonumber\\
&+& 2\Delta \mu\coth \left(\frac{\Delta \mu}{2k_BT}\right)G_0\sum_{i }\tau_i (1-\tau_i).
\label{eq:shot}
\eea
Here, $T=T_{L,R}$ and $\Delta \mu = \mu_L-\mu_R= eV$ 
is the chemical potential difference due to the bias voltage $V$,
$G_0=2e^2/h$ is the quantum of conductance. The current and the 
power noise include contributions from several channels, with $\tau_i$ 
the transmission probability of the $i$th channel evaluated 
at the equilibrium Fermi energy  $\mu=(\mu_L+\mu_R)/2$ of the metal leads; the chemical potential is shifted 
symmetrically between the $L$ and $R$ terminals.
Eq. (\ref{eq:shot}) is well known; we retrieve it in Appendices A-B as a special limit of a more general expression.

Noise measurements in atomic-scale and molecular junctions performed at low bias 
well agree with Eq. (\ref{eq:shot}), see for example Refs.
\cite{Ruitenbeek06,Tal08,Natelson,Scheer,oren13,Vardimon16}. 
Specifically, when the temperature is low relative to the bias, $\coth(\frac{|\Delta \mu|}{2k_B T})\to 1$, and we get
\bea
S_{\Delta \mu}^{T\to 0} = 2|\Delta \mu| G F.
\label{eq:shotV}
\eea
Here,  $F=\sum_i \tau_i(1-\tau_i)/\sum_i \tau_i$ is the Fano factor, $G=G_0\sum_{i}\tau_i$ stands for the electrical conductance 
and $e$ is the charge of the electron.
Equation (\ref{eq:shotV}) is linear in voltage; nonlinearity of the low-temperature high-bias shot noise with voltage is  
referred to as an `anomalous' behavior.
Since $\langle I\rangle=G V$ and $\Delta \mu = eV$,
we can organize Eq. (\ref{eq:shotV}) in its familiar form as $S_{\Delta \mu}^{T\to 0} = 2 e |\langle I \rangle|  F$.

We now consider a junction at equilibrium. Eq. (\ref{eq:S12}) then reduces to the Johnson-Nyquist thermal noise,
\bea
S_{\Delta\mu=0}^T=4k_BT G, 
\label{eq:Seq}
\eea
with the electrical conductance $G=\frac{2e^2}{h}\int d\epsilon \tau(\epsilon)\left(-\frac{df}{d\epsilon}\right)$,
see Appendix A for more details.
Note that we can also approach the equilibrium limit from Eq. (\ref{eq:shot}) and arrive at a corresponding result. Nevertheless,
Eq. (\ref{eq:Seq}) is derived without assuming a constant transmission function.


Fluctuations are enhanced beyond the equilibrium value when a temperature difference is applied across the junction,
as was recently demonstrated in Ref. \cite{Oren}. In this case,
we return to Eq. (\ref{eq:Landauer}) and consider noise contributions due to the temperature difference $\Delta T=T_L-T_R$, 
instead of a voltage difference. Assuming a constant transmission probability we find that 
the electric current vanishes, and that the power noise Eq. (\ref{eq:S12}) simplifies to
\bea
S_{\Delta T}&=&4k_B\bar {T}G_0\sum_{i}\tau_i 
\nonumber\\
&+& 
\left[\frac{k_B(\Delta T)^2}{\bar T} \left(\frac{\pi^2}{9}
-\frac{2}{3}\right) \right] G_0\sum_{i}\tau_i (1-\tau_i).
\label{eq:dtN}
\eea
Here, $\bar T=(T_L+T_R)/2$ is the averaged temperature. 
This result was derived in Ref. \cite{Oren}, see also Appendix C.
The excess (nonequilibrium) delta-T noise, which is the second term in Eq. (\ref{eq:dtN}) displays three particular characteristics: 
(i) It is quadratic in the temperature difference, (ii) it is inversely proportional to the average temperature, 
and (iii) as a partition noise, it is proportional to the factor $\sum_{i}\tau_i (1-\tau_i)$. 

%

In what follows, we ask the following question: How do we  generalize
Eqs. (\ref{eq:shot}) and (\ref{eq:dtN})  
to junctions with an energy dependent transmission function, termed here `resonant tunneling junctions'?
The variation of the transmission function with energy should be taken into account in atomic and molecular junctions
when the voltage bias or the temperature differences are large
such that contributions of carriers beyond the Fermi energy become substantial, see Fig. \ref{Fig0}.


\section{Anomalous noise far from equilibrium} 
\label{sec-mainR}

The transmission function portrays the conducting junction: it is peaked 
at energies of charge conducting atomic/molecular orbitals (resonances) 
and its width reflects the hybridization energy of that state to the metal
electrodes \cite{lathaM}. 
The approximation of a constant transmission probability is meaningful only close enough to equilibrium,
when the width of the transmission function 
(dictated by the coupling energy of the atomic chain or molecule to the electrodes) 
is broad relative to the bias window.
%
Without committing to a particular form, we write down a Taylor expansion 
for the transmission function, performed around the equilibrium Fermi energy $\mu$,
\bea
\displaystyle{\tau(\epsilon)\approx \tau(\mu)+\frac{d\tau}{d\epsilon}\bigg|_{\mu}(\epsilon-\mu)}. 
\label{eq:trane}
\eea
%
For simplicity, we denote $\tau(\mu)$ by $\tau_0$ 
and $\tau'(\mu)\equiv \frac{d\tau}{d\epsilon}\big|_{\mu}$.
As a specific example, consider a resonant level with a single orbital at energy $\epsilon_d$ and 
broadening $\Gamma_{L,R}$ from the left and right metals. The transmission function is given by
\bea
\tau(\epsilon)=\frac{\Gamma_L\Gamma_R}{(\epsilon-\epsilon_d)^2+(\Gamma_L+\Gamma_R)^2/4}.
\label{eq:loren}
\eea
Assuming a broad resonance shifted from the Fermi energy, 
the Lorentzian can be approximated by Eq. (\ref{eq:trane}) with
\bea
\tau_0&=& \frac{\Gamma_L\Gamma_R}{(\epsilon_d-\mu)^2+(\Gamma_L+\Gamma_R)^2/4},\,\,\,\
\nonumber\\
%
\tau'(\mu)&=&\frac{2(\epsilon_d-\mu)\Gamma_L\Gamma_R}{[(\mu-\epsilon_d)^2+(\Gamma_L+\Gamma_R)^2/4]^2}.
\eea
%
Our analytical work does not assume the Lorentzian lineshape for the transmission function;
numerical simulations reported in Sec. \ref{sec-simul} adopt this form.

We now present the main results of our work. We substitute Eq. (\ref{eq:trane})
into Eq. (\ref{eq:S12}), evaluate integrals, and simplify the result in different limits. 
For simplicity, we consider only a single transport channel. 

\vspace{3mm}
\subsection{Voltage-activated noise}

We derive a closed-form expression for the voltage-activated noise at nonzero temperature $T$. 
Details are given in Appendix B. 
Our expression can be organized in various ways to highlight its different properties.
Here, we write it down in a manner that features its dependence on the scaled voltage, defined as
$4k_B TG_0\tau_0  \left[\frac{\Delta \mu}{2k_B T}\coth\left(\frac{\Delta \mu}{2 k_B T}\right)-1 \right]$
(see e.g. \cite{Natelson14}),
\begin{widetext}
\bea
S_{\Delta \mu}^T&=& 
4k_B{T}\tau_0G_0  
\nonumber\\
&+&
4k_B TG_0 \left[\frac{\Delta \mu}{2k_BT}\coth\left(\frac{\Delta \mu}{2 k_B {T}}\right)-1 \right]  \tau_0(1-\tau_0)  
\nonumber\\
&-&
4k_B {T}G_0 \left[\frac{\Delta \mu}{2k_BT}\coth\left(\frac{\Delta \mu}{2 k_B {T}}\right)-1 \right] 
\left[ [\tau'(\mu)]^2 \frac{\pi^2k_B^2{T}^2}{3}  +  [\tau'(\mu)]^2 \frac{(\Delta \mu)^2}{12}     \right]
\nonumber\\
&+& \frac{2}{3} k_B {T}G_0    [\tau'(\mu)]^2  (\Delta \mu)^2.
%
\label{eq:SmuN}
\eea
\end{widetext}
Equation  (\ref{eq:SmuN}) is a central result of our work. We refer to it as the
`voltage activated resonant tunneling noise', given its dependence on the
structure of the resonance, $\tau'(\mu)$. 

The first line in Eq. (\ref{eq:SmuN}) is the Johnson-Nyquist equilibrium noise; all other terms are `excess' noise contributions.
The second line contains the standard, `normal' shot noise expression, which is based on a constant transmission.
The last two lines depict `anomalous' terms: The third line illustrates that 
when studying the excess noise as a function of the scaled bias, 
we should expect a linear dependence at low voltage, but deviations from linearity 
as we depart from equilibrium. Raising the temperature  further
emphasizes deviations from linear behavior with respect to the scaled voltage.  
The last term (fourth line) additionally shifts-enhances the power noise as we depart from equilibrium by applied voltage.

Overall, based on Eq. (\ref{eq:SmuN}) we conclude that:
(i) Excess current noise is a concave function of the scaled voltage, in agreement with
experimental observations  \cite{Natelson14,Tewari}.  
However, (ii) Eq. (\ref{eq:SmuN}) predicts the suppression of the excess noise at high voltage 
relative to the constant $\tau$ expression, rather than an enhancement, 
as observed in e.g. Ref. \cite{Natelson14}.  
Nevertheless, we point out that when other factors are at play (essentially, many-body interactions 
and the opening of additional channels at high voltage), the transmission function can effectively increase with voltage. 
Combining this effect with the concave functional form would result in an overall enhancement of the
noise---on top of a downward curved function.

Experimentally, one could assess this expression by first 
studying the behavior of the noise  at low bias as a function of the scaled voltage,
to extract the linear trend and the associated $\tau_0$ value. Then, subtracting this contribution from the
noise at high bias one could test the applicability of the nonlinear terms [last two lines in Eq. (\ref{eq:SmuN})]  to the specific case.


Equation (\ref{eq:SmuN}) allows us to quickly retrieve the equilibrium noise, given by
\bea
S_{\Delta \mu=0 }^T= 4k_BT G_0\tau_0,
\label{eq:STN}
\eea
which is identical to the case of a constant transmission function. This is expected since
Eq. (\ref{eq:Seq}) is correct in general for an arbitrary $\tau(\epsilon)$.

We now examine the low temperature limit of Eq. (\ref{eq:SmuN}), $\Delta \mu \gg 2k_B T$.
We obtain the following anomalous, nonlinear shot noise,
\bea
S_{\Delta \mu}^{T\to 0}&=&  2G_0\tau_0(1-\tau_0)|\Delta \mu| - G_0 [\tau'(\mu)]^2 \frac{|(\Delta \mu)^3|}{6}. 
\label{eq:shotT0}
\eea
This expression is one of the central results of our work. 
It exposes a cubic dependence of the power noise on bias voltage, on top of the regular linear behavior
[compare Eq. (\ref{eq:shotT0}) to Eq. (\ref{eq:shotV})]. 
It assumes low temperature, $k_BT\ll \Delta \mu$, but under this assumption 
the result is exact for arbitrarily large bias---to the $\tau'(\mu)$ order considered.


We can also Taylor-expand Eq. (\ref{eq:SmuN}) in bias voltage at nonzero temperature.
We observe a quadratic noise-voltage behavior even when taking into account the structure of the transmission function,
\bea
S_{\Delta \mu  \ll T}^T&=&4k_B{T}\tau_0G_0 
+ 2G_0\tau_0(1-\tau_0) \frac{\Delta\mu^2}{6k_B T}  
\nonumber\\
&-&
2G_0[\tau'(\mu)]^2 \frac{\Delta \mu^2}{3}k_B T \left(\frac{\pi^2}{6}-1\right).
\label{eq:SsmuN}
\eea
Unlike Eq. (\ref{eq:shotT0}), this formula is only valid close to equilibrium. 

Overall, our results in this section demonstrate that including the resonance structure leads to noise suppression.
In Figs. \ref{Fig1}-\ref{Fig3}  we assess the accuracy of Eq. (\ref{eq:SmuN}), as well as the 
impact of sharp resonances on the noise.

\subsection{Delta-T noise}

Delta-T noise is activated by temperature differences as implied by its names; measurements
display a quadratic behavior of the noise with the temperature difference \cite{Oren}.
Indeed, the theoretical analysis of Ref. \cite{Oren} confirmed this trend for a constant transmission function, 
see Eq. (\ref{eq:dtN}).
However, far from equilibrium, that is at large temperature differences, 
the structure of the transmission function (resonance)
should be taken into account in the noise calculation, which is what we set to do here.

Substituting Eq. (\ref{eq:trane}) into Eq. (\ref{eq:S12}) we collect quadratic $(\Delta T)^2$, and quartic 
$(\Delta T)^4$
contributions to the delta-T noise, see Appendices C and D. 
We find that quartic  terms are order of magnitude smaller than the quadratic delta-T noise---even
in the most extreme nonequilibrium case of $\Delta T=2\bar{T}$. 
The full expression to the quartic order is quite cumbersome, and here we write it down as
%
\bea
S_{\Delta T}&=& 4k_B\bar T \tau_0 G_0 + G_0k_B\tau_0(1-\tau_0)\left[  a_1 \frac{(\Delta T)^2}{\bar{T}^2}  
-a_2\frac{(\Delta T)^4}{\bar{T^4}} \right]
\nonumber \\
&-&G_0[\tau'(\mu)]^2k_B^3\bar{T}^3  \left[  b_1 \frac{(\Delta T)^2}{\bar{T}^2}  - b_2 \frac{(\Delta T)^4}{\bar{T}^4}
\right].
\label{eq:SdelT4s}
\eea
The coefficients $a_{1,2}$ and $b_{1,2}$ are  given in  Appendix D, and they satisfy
$a_2/a_1\sim 1/40$ and $b_2/b_1\sim  0.1$.
Therefore, for the delta-T noise there is no analog to
 the anomalous term  $[\tau'(\mu)]^2 |(\Delta \mu)^3|$ of equation (\ref{eq:shotT0}).
Here, in contrast, the quadratic order persists even far from equilibrium 
when the transmission resonance is taken into account.
Disregarding the quartic terms we get (see Appendix D),
\bea
S_{\Delta T}&=&
4k_B\bar T \tau_0 G_0+ G_0k_B\tau_0 (1-\tau_0)\frac{(\Delta T)^2}{\bar{T}}\left(\frac{\pi^2}{9}-\frac{2}{3}\right)
\nonumber\\
&-&G_0[\tau'(\mu)]^2  \left(\frac{7\pi^4}{45}-\frac{4\pi^2}{3}\right)k_B^3\bar{T} (\Delta T)^2.
\label{eq:SdTN}
\eea
%
%
Equation (\ref{eq:SdTN}), the `delta-T resonant tunneling noise' generalizes Eq. (\ref{eq:dtN}).
It displays three central characteristics:
(i) It is quadratic in $\Delta T$. (ii)  The third (new) contribution reduces the noise, compared to the case with
a constant transmission. 
(iii) The third term grows with the average temperature, unlike the second term.
(iv) As mentioned above, the quadratic $(\Delta T)^2$ dependence is robust;
 the contribution of the quartic term in the 
delta-T noise is very small, even when taking into account
the energy dependence of the transmission function.
This result is demonstrated in Fig. \ref{Fig4}, which is discussed in the next section.


\section{Simulations}
\label{sec-simul}

We use simulations to assess the validity of our analytic expressions from Sec. \ref{sec-mainR}. 
Assuming coherent tunneling through a single level as described by Eq. (\ref{eq:loren}), the authors of
Ref. \cite{Nitzan06} demonstrated a rich behavior of the voltage-activated current noise as a function of gate voltage.
Here, we present simulations with three objectives in mind. 
First, assuming the Fermi energy is placed on the shoulder of a broad resonance
we exemplify the usefulness of the analytic results of Sec. \ref{sec-mainR},
and relate simulations to experimental observations.
Second, we perform simulations for narrow resonances, 
which are not captured by our theory. Here, we illustrate a significant nonlinearity of
the power noise with voltage far from equilibrium (at low temperature), 
reproducing some of the observations reported in Ref. \cite{Tewari}. 
Lastly, we demonstrate that while the voltage-activated resonant tunneling noise displays `anomalous'
characteristics for narrow resonances at high voltage, 
the delta-T resonant tunneling noise shows a robust quadratic behavior even far from equilibrium.
%

In simulations we adopt a Lorentzian function, Eq. (\ref{eq:loren}), and assume spatial symmetry, 
$\Gamma=\Gamma_{L,R}$.
Numerical results based on Eq. (\ref{eq:S12}) are compared to the 
`normal' constant transmission expressions of Sec. \ref{secT} and to the `resonant tunneling' results of Sec. 
\ref{sec-mainR}.


\begin{figure}[htbp]
\includegraphics[scale=0.37]{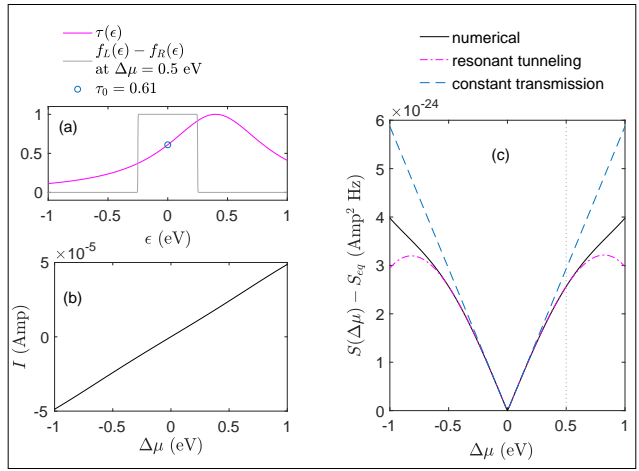}
\caption{
Current and current noise in a tunnel junction with a broad resonance.
(a) Lorentzian transmission function located at $\epsilon_d=0.4$ eV with $\Gamma=0.5$ eV. 
We further present the Fermi function window at $\Delta \mu=0.5$ 
and mark the transmission value at the Fermi energy, $\tau_0=0.61$.
(b) Current-voltage characteristics using the transmission function from panel a.
(c) Excess current noise as a function of voltage:
(full) Numerical calculations with a Lorentzian function using Eq.  (\ref{eq:S12}).
(dashed) Constant transmission expression, Eq. (\ref{eq:shot}) with $\tau_0=0.61$.
(dashed-dotted) Resonant-tunneling result, Eq. (\ref{eq:SmuN}).
The light dotted line identifies the bias $\Delta \mu=0.5$ eV, corresponding to the bias window
in panel (a). 
Simulations were performed at $T=10$ K.}
\label{Fig1}
\end{figure}
\begin{figure}[htbp]
\includegraphics[scale=0.45]{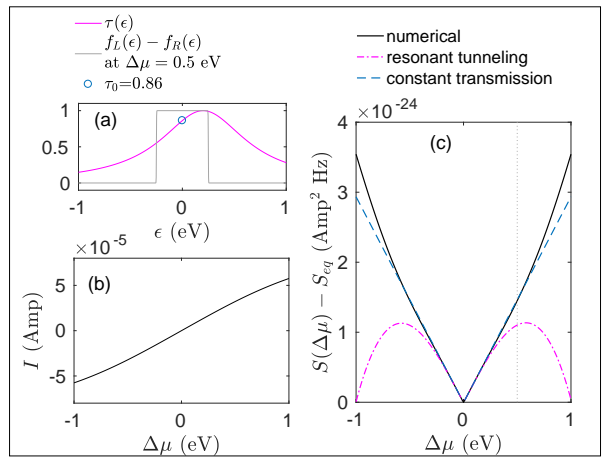}
\caption{Current and current noise as in Fig. \ref{Fig1}, but with  $\epsilon_d=0.2$ eV,  $\Gamma=0.5$ eV.
(a) Lorentzian transmission function and an example of the bias window.
(b) Current-voltage characteristics. 
(c) Current noise as a function of bias voltage. 
The light dotted line identifies the value $\Delta \mu=0.5$ eV, corresponding to the bias window
in panel (a). 
Simulations were performed at $T=10$ K.}
\label{Fig2}
\end{figure}


\begin{figure}[htbp]
\includegraphics[scale=0.38]{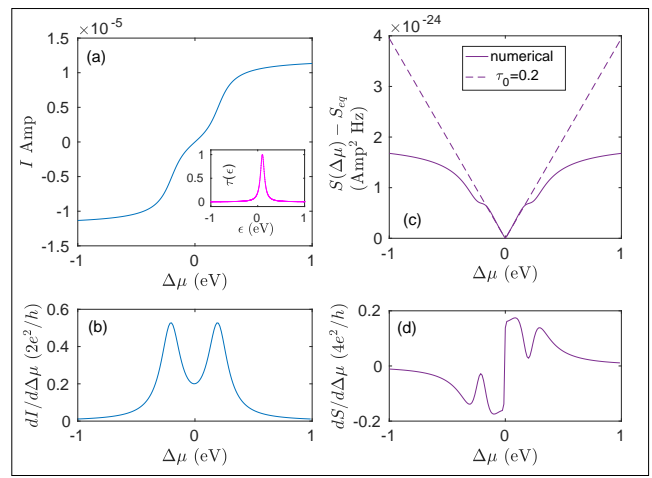}
\caption{
Current and current noise for a junction with a narrow transmission function,
$\epsilon_d=0.1$ eV, $\Gamma=0.05$ eV.
(a) Current-voltage characteristics and the Lorentzian transmission function (inset).
(b) Differential conductance.
(c) Excess noise as a function of voltage bias:
(full) simulations with Eq. (\ref{eq:S12}) are
compared to (dashed) the constant transmission expression 
 Eq. (\ref{eq:shotV}).
(d) Differential excess noise exhibiting nonlinear trends.
Simulations were performed at $T=10$ K.}
\label{Fig3}
\end{figure}
\subsection{Voltage-activated noise}

Roughly, given the low-order Taylor expansion in  Eq. (\ref{eq:trane}), Eq. (\ref{eq:SmuN}) is valid as long as 
$|\tau'(\mu)(\Delta \mu)|$ is smaller than  $\tau_0$,
which is the case  when the Fermi energy is placed on the shoulder of a broad resonance,
 $|\Delta \mu| <2|\epsilon_d-\mu|$ and $\Gamma_{L,R}>\Delta \mu$.


In Fig. \ref{Fig1} we present the current and the excess noise at low temperature 
using a broad transmission function centered at $\epsilon_d=0.4$ eV; 
the Fermi energy is set at zero. We expect Eq. (\ref{eq:SmuN}) to hold 
as long as $\Delta \mu<0.8$, i.e. before the bias window covers the peak of the resonance.
The current-voltage characteristics displayed in panel (b) is quite linear in the full range, which is expected
when using the expansion $\tau(\epsilon)=\tau_0+\tau'(\mu)(\epsilon-\mu)$,
$ \langle I\rangle= \tau_0 G_0 \Delta \mu$.
The current noise is linear in bias close to equilibrium; note that the temperature is quite low.
However, at high bias the current noise as computed numerically from Eq. (\ref{eq:S12}) clearly deviates from 
the linear trend predicted by Eq. (\ref{eq:shotV}). In contrast, the resonant tunneling formula, Eq.
(\ref{eq:SmuN}), provides an excellent match up to $\Delta \mu =0.65$ eV.
This agreement is substantial given that at this point the bias window covers a
large portion of the transmission function, see Fig. \ref{Fig1}(a).
Simulations were performed at 10 K, but similar results were observed for $T$=0.1-300 K.

As mentioned above, Eqs. (\ref{eq:SmuN})-(\ref{eq:shotT0}) should be used with care: They are 
invalid once the voltage window extends over the peak of the resonance. 
Specifically, close to the extremum ($\tau(\epsilon)$=1 here), the low order, linear
expansion (\ref{eq:trane}) is obviously insufficient. 
Thus, when  $\tau_0$ is large, our analytic results hold only at low bias voltage.
%
This point is illustrated in Fig. \ref{Fig2}
with $\epsilon_d$  placed close to the Fermi energy, at $\epsilon_d=0.2$ eV.
We observe in panel (c) that Eq. (\ref{eq:shotT0})  is credible  only for  $\Delta \mu< 0.35$ eV.
%
%
Beyond that, Eq. (\ref{eq:SmuN}) critically fails and in fact (incidentally) 
the constant transmission approximation provides a more accurate result.

It is clear that our results, which are based on
a low-order expansion (\ref{eq:trane}), do not hold if
the resonance is narrow compared to the bias voltage.
In Fig. \ref{Fig3} we study this situation, resorting to direct numerical evaluation of Eq. (\ref{eq:S12}).
As we show in panel (c),  the excess current noise increases linearly at low bias, and it displays a kink 
around $\Delta \mu=0.2$ eV followed by a further increase of noise. 
Overall, the noise is concave in voltage and it is nonlinear, or `anomalous'. 
At the same time, the  differential conductance is symmetric in voltage, see panel (b). 
These concurrent characteristics for the conductance and the noise were reported in Ref. \cite{Tewari}.
Our results quantitatively reproduce these measurements; by further modifying $\epsilon_d$ 
and $\Gamma$ a precise agreement with Ref. \cite{Tewari} can be reached.

It should be pointed out that along with measurements that are reproduced here in Fig. \ref{Fig3}, 
other junctions examined in Ref. \cite{Tewari} displayed more compound, anomalous trends, which are not predicted
by our expressions. These results were explained in Ref. \cite{Tewari} based on interference effects with impurities
at the metal electrodes. 
By further complicating our theory to e.g. emulate many-body interactions, achieved 
by turning $\tau(\epsilon)$ into a function of voltage
and temperature, it should be possible to reproduce more complex noise trends. 
We emphasize however that our objective here has been to stay with
the simple and rigorous coherent transport model and analytically resolve nonlinear characteristics.



\subsection{Delta-T noise}

\begin{figure}[htbp]
\includegraphics[scale=0.38]{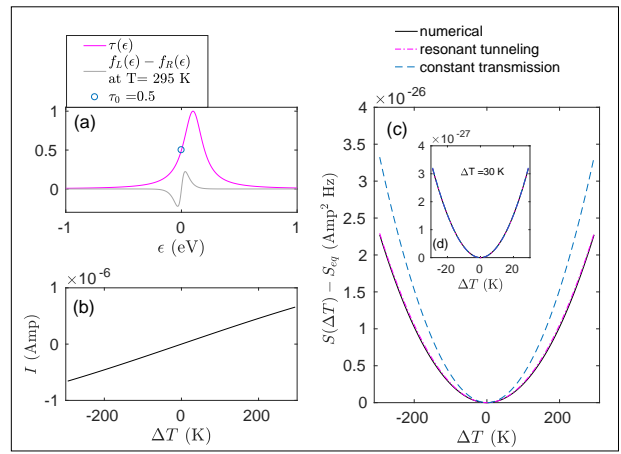}
\caption{
Delta-T noise in resonance tunneling junctions.
(a) Lorentzian transmission function and the delta-T window for $\bar T=300$ K.
(b) Current-$\Delta T$ characteristics at $\bar T=300 $ K.
(c) Excess noise as a function of $\Delta T$  for $\bar T$=300 K and
(d) excess noise at $\Delta T$ =30 K.
Calculations are based on Eq. (\ref{eq:S12}) (full), 
constant transmission formula (\ref{eq:dtN})  (dashed),
and the resonant-tunneling formula (\ref{eq:SdTN}) (dashed-dotted).
Simulations were performed with $\epsilon_d=0.1$ eV, $\Gamma=0.1$ eV,  $\bar T=300$ K, unless otherwise specified.}
\label{Fig4}
\end{figure}

We study the delta-T noise, that is shot noise generated by $\Delta T$ in Fig. \ref{Fig4}. 
As explained in Sec. \ref{sec-mainR}, 
the quadratic formula is robust---as long as the expansion (\ref{eq:trane}) is valid.
Fig. \ref{Fig4} confirms that 
(i) Eq. (\ref{eq:SdTN}) provides a very good approximation to the exact (numerically computed) noise.
(ii) Far from equilibrium, the constant transmission formula Eq. (\ref{eq:dtN}) overestimates the excess noise, 
yet it properly predicts the quadratic behavior.
We further present in panel (d) the excess noise at lower temperature, 
$\bar T$= 30 K, which corresponds to experimental studies \cite{Oren}.
In this case we confirm that the constant transmission expression (\ref{eq:dtN}) is highly accurate.

It is useful to note that under the strict assumption of a constant transmission, the electric current vanishes;
charge current showing up in panel (c) emerges from the contribution of  $\tau'(\mu)$,
\bea
\langle I\rangle&=& \frac{2e}{h}\int_{-\infty}^{\infty} \tau(\epsilon)
\left[ \frac{1}{e^{\beta_L(\epsilon-\mu)}+1}  -  \frac{1}{e^{\beta_R(\epsilon-\mu)}+1} \right]d\epsilon
\nonumber\\
&=&
\frac{2e}{h}
\tau'(\mu)
\int_{-\infty}^{\infty} (\epsilon-\mu)  \left[ \frac{1}{e^{\beta_L(\epsilon-\mu)}+1}  -  \frac{1}{e^{\beta_R(\epsilon-\mu)}+1} \right]d\epsilon
\nonumber\\
&=& \frac{2e}{h}
\tau'(\mu) \frac{\pi^2k_B^2\bar T}{3} \Delta T.
\eea
This delta-T charge current becomes substantial at high temperatures; when $\bar T=30$ K, the electric current 
reaches at most $10^{-8}$A, which is two orders of magnitude smaller than the value at $\bar T=300$ K.

\section{Summary}
\label{sec-summ}

It is often assumed that deviations from the standard 
shot noise formula, Eq. (\ref{eq:shot}) indicate on the
involvement of complex effects beyond those accounted for in the quantum coherent picture, 
e.g. electron heating or electron-phonon interactions.
In contrast, our work here explicates 
that quantum coherent tunneling junctions can support 
`anomalous' shot noise, which should be assessed before
considering additional complex effects.

Revisiting the problem of quantum coherent transport in tunneling junctions,
we derived expressions for the zero frequency current noise under large voltage bias or under 
a temperature difference. 
In such far-from-equilibrium situations, the transmission function
cannot be assumed to be a constant, and one should take into account 
its variation with energy within the nonequilibrium window.
Our main results are that the voltage-activated excess noise is a 
concave function of the scaled voltage, Eq. (\ref{eq:SmuN}), and that
it develops a cubic contribution at low temperature and high voltage, Eq. (\ref{eq:shotT0}). 
We further demonstrated with simulations a highly nonlinear noise behavior, which was in accord
with experimental observations, Fig. \ref{Fig3}.
Finally, we showed that the delta-T noise was well approximated by a quadratic formula, 
$S_{\Delta T}\propto (\Delta T)^2$,
even when the resonance structure was accounted for. 
Our results further manifest that high voltage shot noise measurements can be used to uncover the
structure of the transmission function at the Fermi energy.

We reiterate that quantum coherent transport can support rich phenomena far from equilibrium. 
For example, deviations from the so-called thermodynamic uncertainty relation \cite{udo},
a bound relating dissipation to accuracy (fluctuations), were demonstrated in Ref. \cite{bijayTUR} 
within quantum coherent nanojunctions by considering situations that deviate from Markovianity. 

Beyond the quantum coherent limit, electron correlations, electron-phonon interactions
and other mechanisms for electron scattering should contribute to the appearance of anomalous noise as suggested in Refs. 
\cite{Tewari, Natelson14,Natelson16}, and other studies.
The explorations of such effects in single-molecule junctions, 
to reveal structural and dynamical information is left to future work.

Beyond shot noise, which is the second cumulant of charge fluctuations in steady state, 
a full counting statistics analysis recovers all cumulants, 
and it is useful for characterizing the transport process \cite{FCS1,FCS2}. 
In the context of molecular electronic junctions,
understanding the information concealed in high order moments of the current 
\cite{Reulet,Reznikov,Buttiker11} is left to future work.
 


\begin{acknowledgments}
DS acknowledges fruitful discussions with Oren Tal and funding from the 
Natural Sciences and Engineering Research Council (NSERC) of Canada Discovery Grant 
and the Canada Chairs Program.
\end{acknowledgments}

\begin{widetext}

\renewcommand{\theequation}{A\arabic{equation}}
\setcounter{equation}{0}  
\section*{Appendix A: Derivation of the equilibrium noise Eq. (\ref{eq:Seq})}
\label{AppA}

At equilibrium, $T_L=T_R$, $\mu_L=\mu_R$, Eq. (\ref{eq:S12}) reduces to
\bea
S_1+S_2
=\frac{4e^2}{h}\int_{-\infty}^{\infty}d\epsilon\left(-2k_BT\frac{\partial f}{\partial \epsilon}\right) \tau(\epsilon).
\eea
We identify this contribution by $S_{\Delta\mu=0}^T$, and it corresponds to the Johnson-Nyquist noise,
\bea
S_{\Delta\mu=0}^T= 4k_BTG,
\eea
generated by the thermal fluctuations of charge carriers in conductors at equilibrium.
The linear response conductance is given by
$G= \frac{2e^2}{h}\int d\epsilon \tau(\epsilon) \left(-\frac{df}{d\epsilon}\right)$.

We further separately evaluate $S_1$ in Eq. (\ref{eq:S12}) in the absence of bias voltage,
for general $T_{L}$ and $T_R$. 
These expressions will be of use in Appendix C where we study the delta-T noise.
Using exact identities for the Fermi function we get (omitting the prefactor $4e^2/h$),
\bea
S_1=\int_{-\infty}^{\infty}d\epsilon\left(-k_BT_L\frac{\partial f}{\partial \epsilon}
-k_BT_R\frac{\partial f}{\partial \epsilon}\right)
\left[\tau_0+ \tau'(\mu)(\epsilon-\mu)\right]^2.
\eea
%
We evaluate the different terms and reach
\bea
S_1&=& 
\tau_0^2k_B(T_L+T_R)
\nonumber\\
&+&\int_{-\infty}^{\infty}d\epsilon\left(-k_BT_L\frac{\partial f}{\partial \epsilon}-k_BT_R\frac{\partial f}{\partial \epsilon}\right)2\tau_0 \tau'(\mu) (\epsilon -\mu)
\nonumber \\
&+&\int_{-\infty}^{\infty}d\epsilon\left(-k_BT_L\frac{\partial f}{\partial \epsilon}
-k_BT_R\frac{\partial f}{\partial \epsilon}\right) [\tau'(\mu)]^2 (\epsilon-\mu)^2.
\eea
Using integrals from Appendix E, we put together
\bea
S_1&=2G_0\tau_0^2k_B(T_L+T_R) +2G_0
[\tau'(\mu)]^2 \Big(\frac{\pi^2k_B^3{T_L}^3}{3}
+\frac{\pi^2k_B^3{T_R}^3}{3}\Big),
\label{eq:S1delT}
\eea
where we re-instituted the prefactor $2G_0=4e^2/h$.
We define the averaged temperature,  $\bar T=(T_L+T_R)/2$, and the temperature difference 
$\Delta T= T_L-T_R$. Therefore, $T_{L,R}=(\bar T\pm \Delta T/2)$,  and 
$T_L^3+T_R^3= 2\bar{T}^3 + 3 \bar{T}\Delta T^2/2$.

\renewcommand{\theequation}{B\arabic{equation}}
\setcounter{equation}{0}  
\section*{Appendix B: Derivation of Eq. (\ref{eq:SmuN}): Voltage-activated resonant-tunneling  noise}
\label{AppB}

We examine here the behavior of shot noise under arbitrarily large voltage;  we
assume that there is no applied temperature difference. 
We begin by evaluating $S_1$ in Eq. (\ref{eq:S12}) using the expansion for the transmission function, Eq.
(\ref{eq:trane}).
For convenience, we omit the prefactor $\frac{4e^2}{h}$, and re-install it only at the end of our derivation,
\bea
S_1&=&\int_{-\infty}^{\infty}d\epsilon\left(-k_B{T}\frac{\partial f_L}{\partial \epsilon}
-k_B{T}\frac{\partial f_R}{\partial \epsilon}\right)
\times
\left[\tau_0+ \tau'(\mu)(\epsilon-\mu)\right]^2.
\eea
Here, $f_{\nu}=f(\epsilon, \mu_{\nu}, {T})$,
$\Delta \mu=\mu_L-\mu_R$, $2\mu=\mu_L+\mu_R$ and $T=T_L=T_R$.  
Explicitly,
\bea
S_1&=&\int_{-\infty}^{\infty}d\epsilon\Bigg \{(-k_B {T})\left[\tau_0^2\frac{\partial f_L}{\partial \epsilon}+2\tau_0 \tau'(\mu)(\epsilon-\mu)\frac{\partial f_L}{\partial \epsilon}+[\tau'(\mu)]^2(\epsilon-\mu)^2\frac{\partial f_L}{\partial \epsilon}\right] \nonumber \\
&+&(-k_B {T})\left[\tau_0^2\frac{\partial f_R}{\partial \epsilon}+2\tau_0 \tau'(\mu)(\epsilon-\mu)\frac{\partial f_R}{\partial \epsilon}+[\tau'(\mu)]^2(\epsilon-\mu)^2\frac{\partial f_R}{\partial \epsilon}\right] \Bigg\}.
\eea
We now evaluate the different terms,
\bea
I_1&\equiv&\int_{-\infty}^{\infty}d\epsilon(-k_B {T})\tau_0^2\frac{\partial f_L}{\partial \epsilon}=k_B{T}\tau_0^2, 
\nonumber\\
I_2&\equiv&\int_{-\infty}^{\infty}d\epsilon(-k_B {T})2\tau_0 \tau'(\mu)\left[\epsilon-\left(\mu_L-\frac{\Delta \mu}{2}\right)\right]\frac{\partial f_L}{\partial \epsilon}
\nonumber\\
&=&\int_{-\infty}^{\infty}d\epsilon(-k_B {T})2\tau_0 \tau'(\mu)(\epsilon-\mu_L)\frac{\partial f_L}{\partial \epsilon}
+\int_{-\infty}^{\infty}d\epsilon(-k_B {T})2\tau_0 \tau'(\mu)\frac{\Delta \mu}{2}\frac{\partial f_L}{\partial \epsilon} \nonumber \\
&=&k_B{T}\tau_0 \tau'(\mu)\Delta \mu,
 \nonumber\\
I_3&\equiv& \int_{-\infty}^{\infty}d\epsilon(-k_B {T})[\tau'(\mu)]^2\left[\epsilon-\left(\mu_L-\frac{\Delta \mu}{2}\right)\right]^2\frac{\partial f_L}{\partial \epsilon}
 \nonumber\\
&=&\int_{-\infty}^{\infty}d\epsilon(-k_B {T})[\tau'(\mu)]^2\left[(\epsilon-\mu_L)^2+\Delta \mu (\epsilon-\mu_L)+\frac{(\Delta \mu)^2}{4}\right]\frac{\partial f_L}{\partial \epsilon} 
\nonumber \\
&=&k_B{T}[\tau'(\mu)]^2\frac{\pi^2k_B^2{T}^2}{3}+k_B{T}[\tau(\mu)]^2\frac{(\Delta \mu)^2}{4}.
\eea
%
Summing up these integrals, along with the corresponding contributions from the right side, we get
\bea
S_1=2k_B{T}\tau_0^2+2k_B{T}[\tau'(\mu)]^2\left[\frac{\pi^2k_B^2\bar{T}^2}{3}+\frac{(\Delta \mu)^2}{4}\right].
\label{eq:S1mu}
\eea
Next, we evaluate $S_2$ in Eq. (\ref{eq:S12}). Under bias voltage it can be organized as
\bea
S_2&=&\coth\left(\frac{\Delta \mu}{2 k_B {T}}\right)\int_{-\infty}^{\infty}d\epsilon
\left[f_L(\epsilon)-f_R(\epsilon)\right]
[\tau_0+\tau'(\mu)
(\epsilon-\mu)][1-\tau_0-\tau'(\mu)
(\epsilon-\mu)].
\eea
This integral can be evaluated {\it exactly} using the following relations,
\bea
I_4&\equiv&\coth\left(\frac{\Delta \mu}{2 k_B {T}}\right)\int_{-\infty}^{\infty}d\epsilon[f_L(\epsilon)-f_R(\epsilon)]\tau_0(1-\tau_0)
=\tau_0(1-\tau_0)\Delta \mu\coth\left(\frac{\Delta \mu}{2 k_B {T}}\right),
\nonumber\\
I_5&\equiv&\coth\left(\frac{\Delta \mu}{2 k_B {T}}\right)\int_{-\infty}^{\infty}d\epsilon[f_L(\epsilon)-f_R(\epsilon)](1-2\tau_0)\tau'(\mu)(\epsilon-\mu)=0, 
\nonumber\\
I_6&\equiv&\coth\left(\frac{\Delta \mu}{2 k_B {T}}\right)\int_{-\infty}^{\infty}d\epsilon[f_L(\epsilon)-f_R(\epsilon)][\tau'(\mu)]^2(\epsilon-\mu)^2 \nonumber \\
&=&\coth\left(\frac{\Delta \mu}{2 k_B {T}}\right)[\tau'(\mu)]^2\left[\Delta \mu\frac{\pi^2k_B^2\bar{T^2}}{3}+\frac{1}{12}(\Delta \mu)^3\right].
\eea
Overall, we get
\bea
S_2=\tau_0(1-\tau_0)\Delta \mu\coth\left(\frac{\Delta \mu}{2 k_B {T}}\right)
-\coth\left(\frac{\Delta \mu}{2 k_B {T}}\right)[\tau'(\mu)]^2\left[\Delta \mu\frac{\pi^2k_B^2\bar{T^2}}{3}+\frac{1}{12}(\Delta \mu)^3\right].
\eea
Summing up $S_1$ [Eq. (\ref{eq:S1mu})] and $S_2$, we get the voltage-activated  resonant-tunneling shot noise,
\bea
S_{\Delta\mu}^T&=&2k_B{T}\tau_0^2+2k_B{T}[\tau'(\mu)]^2\left[\frac{\pi^2k_B^2{T}^2}{3}+\frac{(\Delta \mu)^2}{4}\right]
 \nonumber \\
&+&\tau_0(1-\tau_0)\Delta \mu\coth\left(\frac{\Delta \mu}{2 k_B \bar{T}}\right)
-\coth\left(\frac{\Delta \mu}{2 k_B {T}}\right)[\tau'(\mu)]^2\left[\Delta \mu\frac{\pi^2k_B^2{T^2}}{3}+\frac{1}{12}(\Delta \mu)^3\right].
\eea
Multiplying it by $2G_0=\displaystyle{\frac{4e^2}{h}}$ we obtain Eq. (\ref{eq:SmuN}).

\renewcommand{\theequation}{C\arabic{equation}}
\setcounter{equation}{0}  
\section*{Appendix C: Derivation of Eq. (\ref{eq:SdTN}) for the resonant-tunneling delta-T noise}
\label{AppC}

We study here the behavior of the shot noise at finite $\Delta T$, without applying a bias voltage.
For convenience, we omit the prefactor $4e^2/h$. 
 Eq. (\ref{eq:trane}) is used inside Eq. (\ref{eq:S12}), resulting in
\bea
S_2&=&\int_{-\infty}^{\infty}d\epsilon
\left[f(\epsilon, \mu,T_L)-f(\epsilon, \mu, T_R)\right]
\coth\left(\frac{\Delta \beta (\epsilon-\mu)}{2}\right)[\tau_0+\tau'(\mu)
(\epsilon-\mu)][1-\tau_0-\tau'(\mu)
(\epsilon-\mu)]
\nonumber \\
&=&\tau_0(1-\tau_0)k_B\left[(T_L+T_R)+
\frac{(T_L-T_R)^2}{2\bar{T}}\left(\frac{\pi^2}{9}-\frac{2}{3}\right)\right]
\nonumber \\
&+& \int_{-\infty}^{\infty} d\epsilon \left[f(\epsilon, \mu,T_L)-f(\epsilon, \mu, T_R)\right]
\coth\left(\frac{\Delta \beta (\epsilon-\mu)}{2}\right)(1-2\tau_0)\tau'(\mu)
(\epsilon-\mu) 
\nonumber \\
&-&\int_{-\infty}^{\infty}d\epsilon 
\left[f(\epsilon, \mu,T_L)-f(\epsilon, \mu, T_R)\right]
\coth\left(\frac{\Delta \beta (\epsilon-\mu)}{2}\right)[\tau'(\mu)]^2(\epsilon-\mu)^2.
\label{eq:int}
\eea
Here, $\Delta \beta=\beta_R-\beta_L$.
We Taylor-expand the two functions around the equilibrium (average) temperature up to $(\Delta T)^4$. These series will be used in Appendix D as well,
\bea 
f(\epsilon, \mu,T_L)-f(\epsilon, \mu, T_R)&\approx &
\Delta T \frac{\partial f}{\partial \bar T} + \frac{1}{24}(\Delta T)^3\frac{\partial^3 f}{\partial \bar T^3} + 
\frac{1}{16\cdot 5!}(\Delta T)^5\frac{\partial^5 f}{\partial \bar T^5} + ...
\nonumber\\
\coth(\Delta\beta(\epsilon-\mu)/2) &\approx&
\frac{2}{\Delta \beta (\epsilon-\mu)}  + \frac{\Delta \beta (\epsilon-\mu)}{6} - \frac{(\Delta \beta)^3(\epsilon-\mu)^3}{360} +...
\eea
We note that
\bea
\displaystyle{\int_{-\infty}^{\infty}d\epsilon [f(\epsilon, \mu,T_L)-f(\epsilon, \mu, T_R)]\coth
\left(\frac{\Delta \beta (\epsilon-\mu)}{2}\right)(\epsilon-\mu)} = 0,
\eea
since the integrand is a product of three odd functions.
Next, we evaluate the last integral in Eq. (\ref{eq:int}) to the quadratic order in $\Delta T$.
From the Taylor expansion, we collect three contributions.
\bea
&&\int_{-\infty}^{\infty}d\epsilon\frac{2}{\Delta \beta (\epsilon-\mu)}\Delta T \frac{\partial f}{\partial \bar{T}}(\epsilon-\mu)^2
=   2\frac{\pi^2 k_B^3 \bar{T}}{3}\left[\bar{T}^2-\frac{(\Delta T)^2}{4}\right],
\nonumber\\
&&\int_{-\infty}^{\infty}d\epsilon\frac{2}{\Delta \beta (\epsilon-\mu)}\frac{1}{24}(\Delta T)^3\frac{\partial^3 f}{\partial \bar{T}^3}(\epsilon-\mu)^2=0,  
\nonumber\\
&&\int_{-\infty}^{\infty}d\epsilon\frac{\Delta \beta (\epsilon-\mu)}{6}\Delta T \frac{\partial f}{\partial \bar{T}}(\epsilon-\mu)^2
= \frac{7\pi^4}{90}k_B^3\bar{T} (\Delta T)^2 + {\cal O}(\Delta T^4).  
\eea
Note that $\Delta \beta=\frac{\Delta T}{\bar T^2-\Delta T^2/4}$.
Altogether, to the quadratic order we get,
\bea
S_2&=&\tau_0(1-\tau_0)k_B
\left[(T_L+T_R)+\frac{(T_L-T_R)^2}{2\bar{T}}\left(\frac{\pi^2}{9}-\frac{2}{3}\right)\right]
\nonumber \\ 
&-&[\tau'(\mu)]^2 \left[\frac{2\pi^2k_B^3\bar{T}^3}{3}+ \left(\frac{7\pi^4}{90}-\frac{\pi^2}{6}\right)k_B^3\bar{T} (\Delta T)^2 \right].
\label{eq:S2delT}
\eea
We re-install the factor $4e^2/h$ and add Eq. (\ref{eq:S1delT}), 
\bea
S_{\Delta T}&=&
4k_B\bar T \tau_0 G_0+ G_0k_B\tau_0 (1-\tau_0)\frac{(\Delta T)^2}{\bar{T}}\left(\frac{\pi^2}{9}-\frac{2}{3}\right)
\nonumber\\
&-&G_0[\tau'(\mu)]^2  \left(\frac{7\pi^4}{45}-\frac{4\pi^2}{3}\right)k_B^3\bar{T} (\Delta T)^2.
\eea
This expression generalizes the results of Ref. \cite{Oren}. It demonstrates that
even when the transmission function varies with energy,
$\tau'(\mu)\neq0$, the delta-$T$ noise
scales as $\Delta T^2$. Unlike the zero order ($\tau_0$) contribution, the new term takes a negative sign.
As well, the last term scales linearly with the averaged temperature, 
unlike the case of constant transmission.
In Appendix D we furthermore evaluate the contribution of quartic terms $(\Delta T)^4$
to the resonant-tunneling delta-T noise.

\renewcommand{\theequation}{D\arabic{equation}}
\setcounter{equation}{0}  
\section*{Appendix D: Quartic contributions to the resonant tunneling delta-T noise}

Complementing Appendix C, we evaluate here $S_2$ up to fourth order in $\Delta T$.
Some of the integrals were evaluated in Appendix C--- but to the quadratic order.
First, we consider the following integral,
\bea
I_7&\equiv&\int_{-\infty}^{\infty}d\epsilon\frac{2}{\Delta \beta (\epsilon-\mu)}\Delta T \frac{\partial f}{\partial \bar{T}}(\epsilon-\mu)^2
\nonumber\\
&=& 
\frac{2\pi^2 k_{B}^3 \bar{T}}{3}\left[\bar{T}^2-\frac{(\Delta T)^2}{4}\right].
\eea
Recall that $\Delta \beta=\beta_R-\beta_L$ and that $\Delta T=T_L-T_R$, $\bar T=(T_L+T_R)/2$. We further evaluate
\bea
I_8&\equiv&\int_{-\infty}^{\infty}d\epsilon\frac{2}{\Delta \beta (\epsilon-\mu)}\frac{1}{24}(\Delta T)^3\frac{\partial^3 f}{\partial \bar{T}^3}(\epsilon-\mu)^2=
0,
\eea
\bea
I_9&\equiv&\int_{-\infty}^{\infty}d\epsilon\frac{\Delta \beta (\epsilon-\mu)}{6}\Delta T \frac{\partial f}{\partial \bar{T}}(\epsilon-\mu)^2
\nonumber\\
&=&\frac{7\pi^4k_{B}^3\bar{T}^3}{90}\frac{(\Delta T)^2}{\bar{T}^2-\frac{(\Delta T)^2}{4}} 
=\frac{7\pi^4k_{B}^3\bar{T}^3}{90}\frac{(\Delta T)^2}{\bar{T}^2}\left[1+\frac{(\Delta T)^2}{4\bar{T}^2}\right] +{\cal O}(\Delta T^6),
\eea
and
\bea
I_{10}&\equiv&
\int_{-\infty}^{\infty}d\epsilon\frac{\Delta \beta (\epsilon-\mu)}{6}\frac{1}{24}(\Delta T)^3\frac{\partial ^3 f}{\partial \bar{T}^3}(\epsilon-\mu)^2
\nonumber\\
&=&\frac{\Delta \beta}{144}(\Delta T)^3\frac{14\pi^4}{5}k_B^4\bar T 
=\frac{1}{144}\frac{14\pi^4}{5}k_B^3\frac{(\Delta T)^4}{\bar{T}} +{\cal O}(\Delta T^6),
\eea
\bea
I_{11}&\equiv&
\int_{-\infty}^{\infty}d\epsilon\frac{(\Delta \beta)^3(\epsilon-\mu)^3}{360}\Delta T \frac{\partial f}{\partial \bar{T}}(\epsilon-\mu)^2
\nonumber\\
&=&\frac{1}{360}\frac{(\Delta \beta)^3 \Delta T}{\bar{T}}\left(\frac{31}{21}\pi^6k_B^6\bar{T}^6\right) 
=\frac{1}{360}\frac{31}{21}\pi^6k_B^3\frac{(\Delta T)^4}{\bar{T}} +{\cal O}(\Delta T^6),
\eea
\bea
I_{12}&\equiv&
\int_{-\infty}^{\infty}d\epsilon\frac{2}{\Delta \beta (\epsilon-\mu)}\frac{1}{16\times 5!}(\Delta T)^5\frac{\partial^5 f}{\partial \bar{T}^5} (\epsilon-\mu)^2=0.
\eea
Summing up these terms up to fourth order in $\Delta T$ we obtain 
\bea
S_2&=&\tau_0(1-\tau_0)k_B\left\{\left[(T_L+T_R)+
\frac{(\Delta T)^2}{2\bar{T}}\left(\frac{\pi^2}{9}-\frac{2}{3}\right)\right]+\frac{(\Delta T)^4}{\bar{T^3}}\left[\frac{\pi^2}{72}-\frac{1}{24}+\frac{1}{40}-\frac{7}{15}\frac{\pi^4}{360}\right]\right\} 
\nonumber \\
&-&[\tau'(\mu)]^2 \left\{\left[\frac{2\pi^2k_B^3\bar{T}^3}{3}+ \left(\frac{7\pi^4}{90}-\frac{\pi^2}{6}\right)k_B^3\bar{T} (\Delta T)^2 \right]+k_B^3\frac{(\Delta T)^4}{\bar{T}}\left[\frac{7\pi^4}{90}\frac{1}{4}+\frac{1}{144}\frac{14\pi^4}{5}-\frac{1}{360}\frac{31}{21}\pi^6\right]\right\}.
\nonumber \\
\label{eq:S24}
\eea
We add $S_1$ from Eq. (\ref{eq:S1delT}) and furthermore evaluate the coefficients,
\bea
S_{\Delta T}&=& 4k_B\bar T \tau_0 G_0 + G_0k_B\tau_0(1-\tau_0)\left[  0.43 \frac{(\Delta T)^2}{\bar{T}^2}  
-0.012\frac{(\Delta T)^4}{\bar{T^4}} \right]
\nonumber \\
&-&G_0[\tau'(\mu)]^2k_B^3\bar{T}^3  \left[  1.99 \frac{(\Delta T)^2}{\bar{T}^2}  - 0.31 \frac{(\Delta T)^4}{\bar{T}^4}
\right].
\label{eq:SdelT4}
\eea
Only in the most extreme nonequilibrium case of $\Delta T=2\bar{T}$, 
the contribution of the quartic terms is within order of magnitude of the
quadratic term- within the $\tau'$ correction.
Therefore, we can safely ignore quartic contributions to the delta-T noise---as long as 
(\ref{eq:trane}) is valid.
Back to Eq. (\ref{eq:S24}), we combine $S_1$ and $S_2$ up to second order in $\Delta T$,  restore the factor $4e^2/\hbar$, and 
retrieve Eq. (\ref{eq:SdTN}),
\bea
S_{\Delta T}&=&
4k_B\bar T \tau_0 G_0+ G_0k_B\tau_0 (1-\tau_0)\frac{(\Delta T)^2}{\bar{T}}\left(\frac{\pi^2}{9}-\frac{2}{3}\right)
\nonumber\\
&-&G_0[\tau'(\mu)]^2  \left(\frac{7\pi^4}{45}-\frac{4\pi^2}{3}\right)k_B^3\bar{T} (\Delta T)^2.
\eea

\renewcommand{\theequation}{E\arabic{equation}}
\setcounter{equation}{0}  
\section*{Appendix E: Useful integrals}

Consider the Fermi-Dirac function $f(\epsilon)=\left[ e^{\epsilon/T} +1\right]^{-1}$, the following 
relationship and integrals are useful to the evaluation of the shot noise,
\bea
f(\epsilon)\left[ 1-f(\epsilon)\right]= k_BT \left(-\frac{\partial f}{\partial \epsilon}\right),
\,\,\,\,\,\,\
\frac{\partial f}{\partial T}= -\frac{\epsilon}{T}\frac{\partial f}{\partial \epsilon}
\eea
\bea
\int_{-\infty}^{\infty}d\epsilon\left(-\frac{\partial f}{\partial \epsilon}\right)=1,\,\,\,\,\,
\displaystyle{\int_{-\infty}^{\infty}d\epsilon \epsilon \left(-\frac{\partial f}{\partial \epsilon}\right)=0},
\label{eq:int1}
\eea
%
%
\bea
\displaystyle{\int_{-\infty}^{\infty}d\epsilon\epsilon^2
\left(-\frac{\partial f}{\partial \epsilon}\right)=\frac{\pi^2k_B^2T^2}{3}},\,\,\,\,\,\,\,\,\,
\int_{-\infty}^{\infty}d\epsilon\epsilon^4\left(-\frac{\partial f}{\partial \epsilon}\right)= \frac{7\pi^4k_B^4T^4}{15},\,\,\,\,\,\,\,\,\,
\displaystyle{\int_{-\infty}^{\infty}d\epsilon \epsilon^6 \left(-\frac{\partial f}{\partial \epsilon}\right)=\frac{31\pi^6k_B^6T^6}{21}},
\eea
\bea
\int_{-\infty}^{\infty}d\epsilon\epsilon^2\left(-\frac{\partial f}{\partial \epsilon}\right)^2= \frac{k_BT}{2}\left(\frac{\pi^2}{9}-\frac{2}{3}\right),
\eea
\bea
\int_{-\infty}^{\infty}d\epsilon\frac{1}{\epsilon}\left(\frac{\partial^3f}{\partial T^3}\right)= \frac{2}{T^3},\,\,\,\,\,\,\,\,
\int_{-\infty}^{\infty}d\epsilon\frac{1}{\epsilon}\left(\frac{\partial^5f}{\partial T^5}\right)= \frac{24}{T^5},
\eea
\bea
\displaystyle{\int_{-\infty}^{\infty}d\epsilon \epsilon \left(\frac{\partial^3 f}{\partial T^3}\right)=0},\,\,\,\,\,\,\,\,
\displaystyle{\int_{-\infty}^{\infty}d\epsilon \epsilon \left(\frac{\partial^5f}{\partial T^5}\right)=0},
\eea 
\bea
\displaystyle{\int_{-\infty}^{\infty}d\epsilon \epsilon^3 \left(\frac{\partial^3 f}{\partial T^3}\right)=\frac{14\pi^4k_B^4T}{5}}.
\label{eq:int3}
\eea 

\end{widetext}



\begin{thebibliography}{4}
\bibitem{1overf}
Low frequency $1/f$ noise is not considered in the present analysis.
\end{thebibliography}
\end{document}